\documentclass{article}

\usepackage{arxiv}

\usepackage[utf8]{inputenc} 
\usepackage[T1]{fontenc}    
\usepackage{hyperref}       
\usepackage{url}            
\usepackage{booktabs}       
\usepackage{amsfonts}       
\usepackage{nicefrac}       
\usepackage{microtype}      
\usepackage{lipsum}
\usepackage{graphicx}
\graphicspath{ {./images/} }

\title{Matters of Life and Death in Computational Cell Biology}

\author{
 Connor McShaffrey \\
  Cognitive Science Program\\
  Indiana University Bloomington\\
  Bloomington, IN, USA \\
  \texttt{cmcshaff@iu.edu} \\
   \And
 Eran Agmon \\
  Center for Cell Analysis and Modeling\\
  University of Connecticut Health\\
  Storrs, CT, USA \\
  \And
 Randall D. Beer \\
  Cognitive Science Program\\
  Program in Neuroscience \\
  Informatics Department \\
  Indiana University Bloomington\\
  Bloomington, IN, USA \\
}

\begin{document}
\maketitle
\begin{abstract}
Nearly all cell models explicitly or implicitly deal with the biophysical constraints that must be respected for life to persist. Despite this, there is almost no systematicity in how these constraints are implemented, and we lack a principled understanding of how cellular dynamics interact with them and how they originate in actual biology. Computational cell biology will only overcome these concerns once it treats the life-death boundary as a central concept, creating a theory of cellular viability. We lay the foundation for such a development by demonstrating how specific geometric structures can separate regions of qualitatively similar survival outcomes in our models, offering new global organizing principles for cell fate. We also argue that idealized models of emergent individuals offer a tractable way to begin understanding life’s intrinsically generated limits. 
\end{abstract}


\section{Introduction}
One of the most fundamental questions in biology is how to understand what stands between a cell and mere chemical processes: the boundary between life and death. While this concept may seem abstract, its importance is evident in numerous biological problems, ranging from determining habitable environments, to designing treatments that selectively target and kill pathological cells, to proper tissue and organ development. As fundamentally precarious and mortal agents, it is not possible to comprehensively discuss the lives of cells without also confronting their biophysical limits.

Despite the centrality of the life-death issue, we currently lack a principled way of describing what generates this boundary and how a cell’s dynamics unfold relative to it. When in a cell’s decay should it be considered dead, and at which points should it be considered as good as dead without intervention? While biologists usually get by identifying living organisms with an attitude of “I’ll know it when I see it,” numerous cases challenge this intuition. Biological systems are incredibly high-dimensional, and many of the intertangled factors that lead to their death are not accounted for. Cells can persist after events that were previously thought to be points of no return in apoptosis \cite{spencer_measuring_2011, bock_mitochondria_2020}, dormant organisms with nearly nonexistent metabolic activity can be brought back from the brink \cite{lennon_microbial_2011}, and synthetic cells challenge us to think about when an engineered system should be considered alive \cite{elowitz_build_2010}. 

Even within our mathematical and computational models, some of our most powerful tools for constructing rigorous theories in biology, the constraints and deaths of cells are handled in an unsystematic and \textit{ad hoc }manner. Models are constructed by identifying a few key variables and parameters and defining their interrelations, but often the conditions for life and death are not specified with the same level of rigor. Even when models are explicit about their death conditions, there is almost no consistency in the implementation between them, and within a single model the details are usually relegated to the supplementary materials.  This is counterproductive, as the existential features of any model are only captured when we consider the rules of death in unity with those that govern the living dynamics. 

In this paper, we aim to provide a theoretical foundation for addressing cellular viability and death. Section II reviews how the life-death distinction is traditionally handled in the cell modeling literature, focusing on models of cell death and highlighting the assumptions implicit in these formalisms—specifically, the reliance on externally imposed, approximated death functions. Section III demonstrates that a more rigorous theoretical approach to death can be developed by viewing death functions as geometrically delineating the space of habitable states, a \textit{viability region}. Whereas many biological modelers have an intuitive sense of cell fates unfolding in an abstract geometric space \cite{waddington_strategy_1957, ferrell_bistability_2012, lange_cellrank_2022}, with recent single-cell approaches beginning to reconstruct analogous geometric landscapes experimentally \cite{bergen_generalizing_2020,weiler_cellrank_2024}, the viability region is an additional layer of geometric structure that is often ignored. Section IV further develops the geometric perspective by showing how the interactions between the dynamics and viability region results in three new classes of global manifolds which decompose the viability region into subregions according to whether states will result in survival or death. Whereas section IV analyzes the consequences of \textit{extrinsically imposed} death functions, section V interrogates the origin of the viability region as an \textit{intrinsically generated} property of life as an emergent molecular system, and demonstrates some basic principles of how this emergent viability region can be analyzed using idealized theoretical models. Finally, Section VI outlines additional work that needs to be done to develop an understanding of cellular viability that can be applied to various classes of biological models. 

\section{Death in Cell Models}

The death of a cell marks its transition between a living state (where the biology of its physiological processes dominates) and a nonliving state (where the inanimate chemistry of its former components dominates). The ways this boundary can be crossed are varied and can even play roles in other biological processes \cite{ghose_cell_2020}. However, they are divided broadly into two classes. \textit{Unregulated cell death} is what people typically imagine when they think of a fatal event, with a cell’s physiological vulnerabilities being violated by stressors such as starvation, toxins, and predation. Bacteria, for example, can be consumed by predators through a variety of mechanisms, including attachment, enzyme release, or quorum-mediated lysis \cite{perez_bacterial_2016}. Antibiotics exploit these same vulnerabilities, such as ampicillin disrupting cell wall formation by inhibiting penicillin-binding proteins \cite{cho_beta-lactam_2014, wong_understanding_2021}. Similarly, deprivation of essential molecules can result in cells undergoing metabolic collapse and dying via necrosis \cite{barros_necrotic_2001}. In contrast, \textit{regulated cell death} proceeds in a structured fashion that often takes the function of surrounding cells into consideration. For example, apoptosis proceeds through dedicated internal and external pathways, leading to caspase-mediated self-digestion and the formation of apoptotic bodies that prevent immune activation \cite{elmore_apoptosis_2007}. Necroptosis and pyroptosis initiate cell death in a way that helps coordinate transient increases in inflammation so the body can better respond to situations like a pathogen invading the body \cite{bergsbaken_pyroptosis_2009, pasparakis_necroptosis_2015}. Ferroptosis is characterized by the accumulation of oxidized lipids and is tightly linked to the disruption of lipid homeostasis, highlighting how membrane composition and metabolic pathways can actively govern cell fate \cite{agmon_lipid_2017}. Whether it be due to regulated or unregulated death processes, cells can have biphasic responses to common chemicals in their environment, where either too much or too little results in fatality \cite{karin_biphasic_2017}.

There are similarly many approaches to the treatment of death in computational cell models, each offering different perspectives based on the problem at hand. While a real cell is a high-dimensional emergent system and dissipative structure that arises through a concert of energy flow and molecular interactions \cite{sole_fundamental_2024}, it is currently infeasible to build a model at this resolution. To get around this, computational biologists declare a cell’s existence axiomatically, and construct simplified models of reduced dimensions through a combination of coarse-graining and identifying subsystems of interest. Having rendered the cells' living status as a starting assumption, one then draws upon the experimental literature to determine which conditions are most indicative of death, and formulate a corresponding death function for the model. Far from a subordinate detail, this decision has far-reaching consequences for understanding the limits of cellular physiology and multicellular dynamics in the model. Nonetheless, there is very little systematicity in how cell death is implemented across models.

When studying the underlying mechanisms of a cell death process, most models focus on the dynamics of a small subcellular network. Counterintuitively, however, the actual conditions for death are not always defined. This is in part because, while we have strong theoretical tools to describe changes in concentration and gene activation, the actual limits of life are an elusive concept. For example, ODE models of apoptosis and pyroptosis often identify life and death with the presence of attractors, where the latter involves the presence of death indicators, such as a positive concentration of executioner caspases \cite{eissing_bistability_2004, spencer_measuring_2011, yin_cell_2021, li_caspase-1_2022}. But at what point does the value of such a variable actually indicate death? If the death indicator is the percentage of cleaved substrate \cite{rehm_systems_2006}, it is unambiguous that 100\% cleavage means the cell has died, but what of intermediate values? Far from trivial, incomplete commitment to cell death has been linked to both cell recovery and multicellular pathology \cite{bock_mitochondria_2020}. Beyond this, an attractor-based analysis neglects the possibility of cell’s violating their physiological bounds during their transient dynamics, long before any equilibration takes place. Assigning a model an explicit death condition, such as in Schink et al.’s model of \textit{E. coli} under starvation conditions, is essential to capturing how transient activity plays into survival outcomes \cite{schink_survival_2024, xu_electric_2024}.

While differential equation models of subcellular networks permit detailed investigations when we have approximations for the kinetic parameters, limitations become apparent as genome-level reconstructions increase both the system size and the number of unknowns. Flux Balance Analysis makes tradeoffs to overcome these barriers, avoiding dependence on exact kinetic parameters by optimizing an objective function based on the constrained fluxes of a stoichiometric matrix under an assumed steady state \cite{orth_what_2010}. Despite the vast amount of information present in these models, the conditions for any single cell’s death are often even more ambiguous than they were in the differential equation models. This is because FBA models typically emphasize persistence at the level of the \textit{population}, blurring the unicellular and multicellular scales. For example, when biomass is the objective function, cellular persistence can be formulated relative to the reconstructed network’s capacity to counter the rate of dilution in the environment \cite{regimbeau_contribution_2022}. With dilution being an aggregate measure of the probability of biomass loss per unit time, degradation and death are but two contributing factors. Death at the unicellular scale is only implicitly present, although modifications may be made to change this.

Perhaps surprisingly, continuum and agent-based multicellular models are far more consistent about defining explicit death functions. This is because the dynamics of the conglomerate must continue through any single existential event (birth, transformation, or death), necessitating that the update rules include unambiguous death conditions. The specific way death is handled will impact the way the rest of the population’s subsequent dynamics will be constrained. For example, continuum tumor models track whether available oxygen falls beneath a critical threshold to determine necrotic and proliferative regions of the mass \cite{lowengrub_nonlinear_2010, gefen_modeling_2013}. Meanwhile, agent-based models incorporate death functions into their cells in a variety of ways. Most straightforwardly, death is coded as a conditional, instantaneous update where a cell is either immediately gone or initiates an irreversible death process like the early cellular swelling in necrosis \cite{ghaffarizadeh_physicell_2018}. The death indicators for these functions can be something internal to the cell, such as available ATP, an external factor like position in a chemical gradient, or cell adhesion \cite{gatenby_cellular_2007,smallbone_metabolic_2007,anderson_microenvironment_2009}. These functions can be implemented in a deterministic or probabilistic manner, and at times both will be utilized in a single model \cite{gatenby_cellular_2007,smallbone_metabolic_2007,anderson_microenvironment_2009,ghaffarizadeh_physicell_2018}.

Even as agent-based and individual subcellular models allow us to begin tackling a variety of biological phenomena, they are limited in the sense that many problems in biology require us to consider mechanisms that span spatiotemporal scales. To address this, many groups have converged on \textit{multi-scale modeling }as a gold standard for computational biology \cite{walpole_multiscale_2013}. For example, many agent-based modeling frameworks allow the user to upload subcellular networks to cells in the form of an ODE, stochastic, Bayesian, or FBA submodel \cite{swat_multi-scale_2012,starrus_morpheus_2014,ghaffarizadeh_physicell_2018,ruscone_building_2024}. Given the right interface, numerous subcellular models can even be combined despite the mathematical differences in their construction, allowing us to construct “whole-cell” models \cite{karr_whole-cell_2012, agmon_multi-scale_2020,johnson_building_2023}. Importantly, these models are not reconstructions of true emergent cells, but sophisticated approximations that still declare a priori the existence of the cell and its physiological constraints. The death functions in these models can either be inherited from the submodels that compose them, or defined as a new function based on the values of variables across modules. One example of this is the work of Skalnik et al., where colonies of whole-cell \textit{E. coli }models were simulated to test their response to the antibiotics tetracycline and ampicillin \cite{skalnik_whole-colony_2021}. In this model, the \textit{E. coli }had a module for simulating the integrity of their cell wall, and death was determined as a function of whether the size of  a membrane hole passed a critical threshold that would ultimately lead to lysis. One reason this model has an explicit death function is because, as was argued above for the multicellular models, it must specify when a cell dies and what to do with it. In general, whole-cell models of isolated cells do not have an explicit death function, but at times imply them by focusing on dynamics within physiological conditions.

In summary, while death is a central feature of many biological models, ranging from subcellular to tissue scales, the exact conditions for its occurrence are not always defined. Furthermore, even when these physiological limits are defined, their exact implementation is not consistent, resulting in non-uniform definitions across models. In the following sections, we push toward a more rigorous approach that moves the conditions for survival front-and-center as a unifying characteristic in cellular biology, ultimately opening the door for a more structured analysis of the possible existential outcomes in cellular systems. 

\section{Viability Constraints}

Cells exhibit a rich range of states, from healthy to pathological, homeostatic to degrading, and many computational biologists understand this progression to be dynamically unfolding in a high-dimensional space \cite{lange_cellrank_2022}. However, often neglected in this view is the additional geometric structure which delineates these living states through which cells develop from death — for all cells there is a point at which they irreversibly transition from living to merely chemical systems. While death is almost as varied as life itself, with new classifications and nuances always being discovered, all are modes for this general disintegration of the cellular unit. In the previous section, we showed that whether it be through ATP, executioner caspases, oxygen, or mechanical interactions, every cell model that incorporates death must either explicitly or implicitly have an approximation for these points of no return. While the constraints on cellular physiology are typically specified in a piecemeal fashion, we can also think about their union as a geometric boundary in state space, carving out a \textit{viability region} for a cell. 

The geometric perspective goes back to Ashby, who viewed physiology as taking place within the bounds of certain essential variables \cite{ashby_design_1960}. Since then, this perspective has been developed in multiple fields including adaptive behavior \cite{beer_dynamical_1995}, computational biology \cite{voit_systems-theoretical_2009,davis_dynamical_2019}, dynamical systems \cite{mcshaffrey_decomposing_2023}, control theory \cite{aubin_viability_2011}, and artificial life \cite{barandiaran_norm-establishing_2014}. The collective range of essential variables that fall within the boundary is synonymous with the domain where an agent still exists (Fig. 1A), and states outside of this boundary correspond to death of the cell or organism. The structure of the viability region can then be extended over the dimensions of unconstrained variables that do not alter the structure of the life-death boundary, such as location in the environment (Fig. 2B). This is a global perspective on a single cell’s range of habitable states, in Fig. 1 this is illustrated in 2 and 3 dimensions for visual interpretability, but in general the viability boundary is an \(n-1\)-dimensional manifold embedded in the cell’s \(n\)-dimensional state space.  

\begin{figure}
\centering
\includegraphics[width=5in]{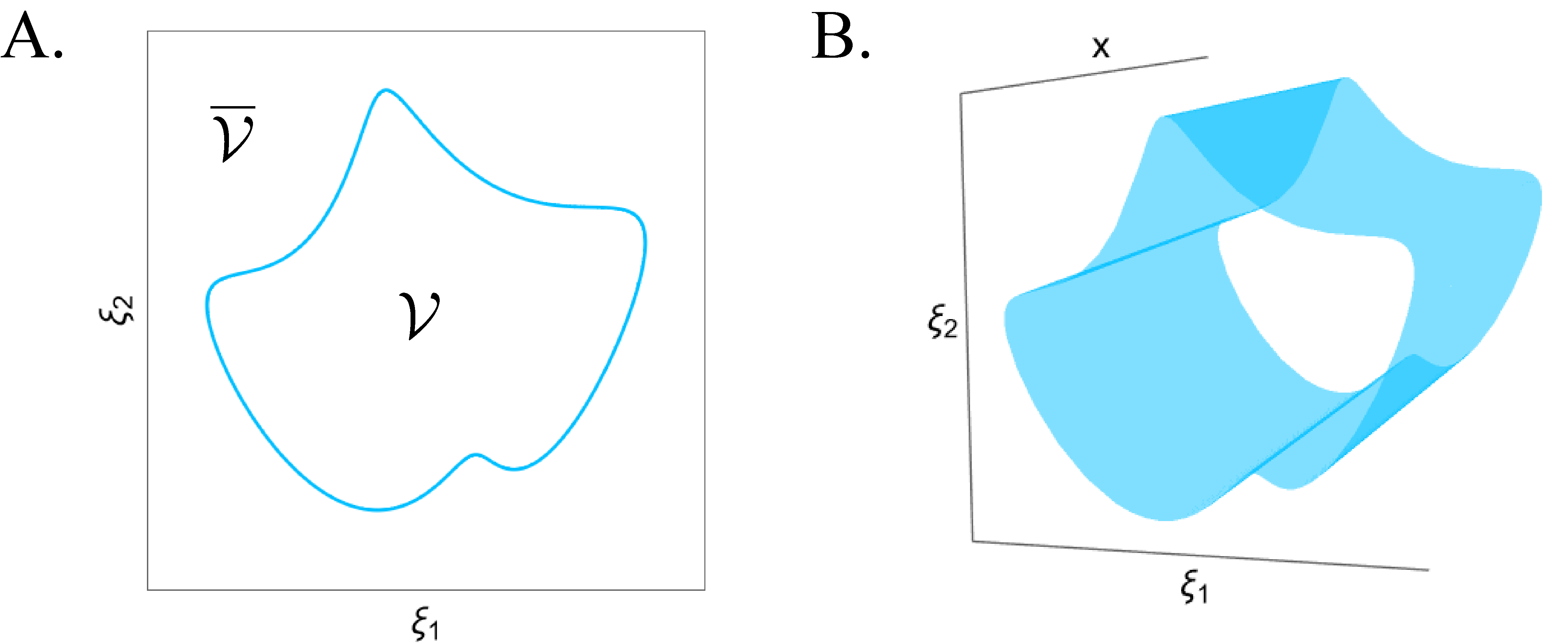}
\caption{{\textbf{Viability space as a geometric concept in a minimal 2D-3D visualization.}}
\textbf{A. }In the space of essential variables, the viability boundary (light blue) marks the edge of the viability region where a cell can still be considered alive. \textbf{B. }The viability region is expanded across the range of unconstrained variables, such as location in the environment, although its structure is not modified across these dimensions.  }
\label{fig1}
\end{figure}

When considering groups of cells acting in aggregate, we need to contend with complexities not present in the unicellular case. Here, death can be a function of spatial interactions as well as internal states \cite{gatenby_cellular_2007,smallbone_metabolic_2007} and the way a cell’s remains are modeled after its death can have a large impact on the subsequent dynamics of cells that remain alive. The simplest implementation is to have the cell mass immediately removed from the simulation following its death \cite{gatenby_cellular_2007, smallbone_metabolic_2007}, but debris such as a calcified body could also be left behind \cite{ghaffarizadeh_physicell_2018,skalnik_whole-colony_2021}. How can we apply the geometric concept of a viability region with this in mind?

While the literature on viability regions almost universally focuses on individual living units, recent work has made progress on the multicellular case by thinking about dynamics as playing out in the intersection of each cell’s viability region \cite{mcshaffrey_dissecting_2024}. Each time a cell enters or leaves the collective, this intersection either grows or collapses by a number of dimensions equal to the number of essential variables belonging to that cell. In Fig. 2, we illustrate a case where each of three coupled cells has an essential variable with an upper and lower bound. When the green cell’s essential variable leaves its viability region, the other cells' dynamics persist in a lower-dimensional subspace. Even if a dying cell leaves behind inert material in the environment, the general idea of the intersection of viability regions collapsing remains intact. As in the single-cell case, the intersection of viability constraints will form an \(n-1\)-dimensional boundary, except now in the collective \(n\)-dimensional state space of the multicellular system. In Fig. 2, this is once again shown in 2 and 3 dimensions for easy visualization. With our viability regions defined in this way, we can begin to consider the global organization of survival outcomes.  

\begin{figure}
\centering
\includegraphics[width=5in]{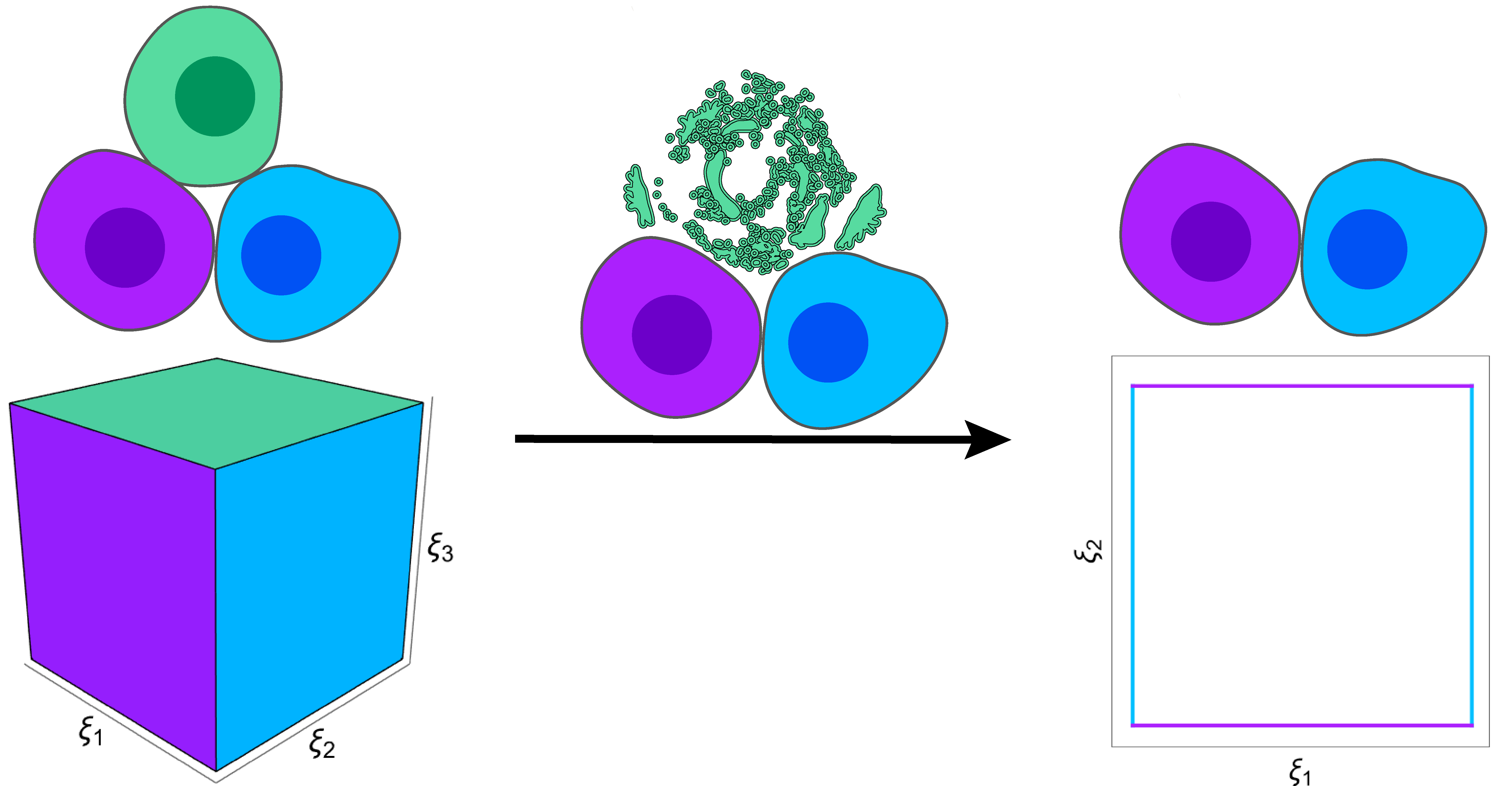}
\caption{\textbf{\textbf{The geometry of multicellular viability constraints.} }
To generalize the geometric idea of life-death boundaries to models of more than one cell, we need to begin thinking about multicellular dynamics as playing out in the intersection of all participating cells’ viability regions. This figure shows an example of three cells’ essential variables in a unified space. When the system’s dynamics result in the green cell violating its viability constraint, it dies, leaving the other two cells behind. This can be imagined as the intersection of viability regions collapsing in its dimensionality.  }
\label{fig2}
\end{figure}

\section{Viability Space Decomposition}

In the previous section, we demonstrated how deterministic death thresholds can be viewed as delimiting a geometric space of physiologically permissible states called a viability region. While deterministic thresholds may be the simplest variety of death function used in computational cell biology, they present an ideal starting point for building new intuitions and are still far from trivial. Recall that even in subcellular ODE models, it is almost never characterized whether a cell can survive the states it will inhabit during its transient dynamics \cite{spencer_measuring_2011}. This is especially concerning in multicellular models, where most of the existential outcomes occur long before the population reaches any kind of steady state. Ideally, we should be able to understand the viability outcomes over short and long timescales without only relying on a large number of simulated trajectories. The aim of this section is to demonstrate that by treating the viability region as a theoretical object of central importance, we can begin deriving organizing principles for cell fate. 

Recently, viability space decomposition has been proposed as one way to understand the global organization of existential outcomes in models governed by nonlinear ordinary differential equations \cite{mcshaffrey_decomposing_2023, mcshaffrey_dissecting_2024}. Building on dynamical systems theory’s approach of identifying the limit sets and separatrices (the manifolds that carve out basins of attraction) which organize the system’s dynamics, this decomposition derives new classes of global manifolds that separate states that will survive from those that will not. This is accomplished by taking advantage of the geometric perspective of a viability region, and thinking carefully about how the dynamics interact with its boundary. In particular, we will introduce \textit{mortality manifolds} (separate surviving and perishing initial conditions), \textit{ordering manifolds} (distinguish initial conditions that will perish by violating different constraints), and \textit{collapse manifolds} (extend mortality manifolds, ordering manifolds, and relevant separatrices to organize multi-agent state spaces). As the mathematical underpinning of these objects are being developed in another paper \cite{mcshaffrey_viability_nodate}, the goal here is simply to stress what a rigorous approach to viability might look like.  

Like the aforementioned analyses of subcellular death circuits, we begin by looking for the presence of a viable attractor, but do not stop here and continue to ask which trajectories can reach this stable state. For example, Fig. 3A shows a case where all the vectors along the viability boundary point toward its interior, meaning that states in the viability region are \textit{asymptotically viable}. This could be an \textit{E. coli }maintaining a homeostatic attractor as it regulates nutrient uptake and metabolism in an environment with reasonable temperature, osmolarity, and glucose availability \cite{kochanowski_global_2021}. Fig. 3B, on the other hand, identifies an attractor (dark blue) outside of the viability region— although the asymptotic state is not reachable, the trajectories approach it in finite time, making the whole region \textit{transiently viable}. For example, when a yeast cell has access to glucose but no other growth-supporting nutrients, it will rapidly produce reactive oxygen species, ultimately leading to its death \cite{granot_sugar-induced_2003}. In each of these cases, the approach of identifying the attractor’s location is sufficient for understanding whether the cell will live or die, but Fig. 3C shows where this breaks down. Here, the attractor is in the viability region, but not all trajectories are capable of reaching it without violating the viability constraints, splitting the habitable states into asymptotically and transiently viable subsets. This case captures the idea of a system attempting to regain homeostasis following a perturbation, but failing to compensate, such as how intermediate DNA damage is sometimes recoverable and other times fatal \cite{ciccia_dna_2010}. 

\begin{figure}
\centering
\includegraphics[width=6in]{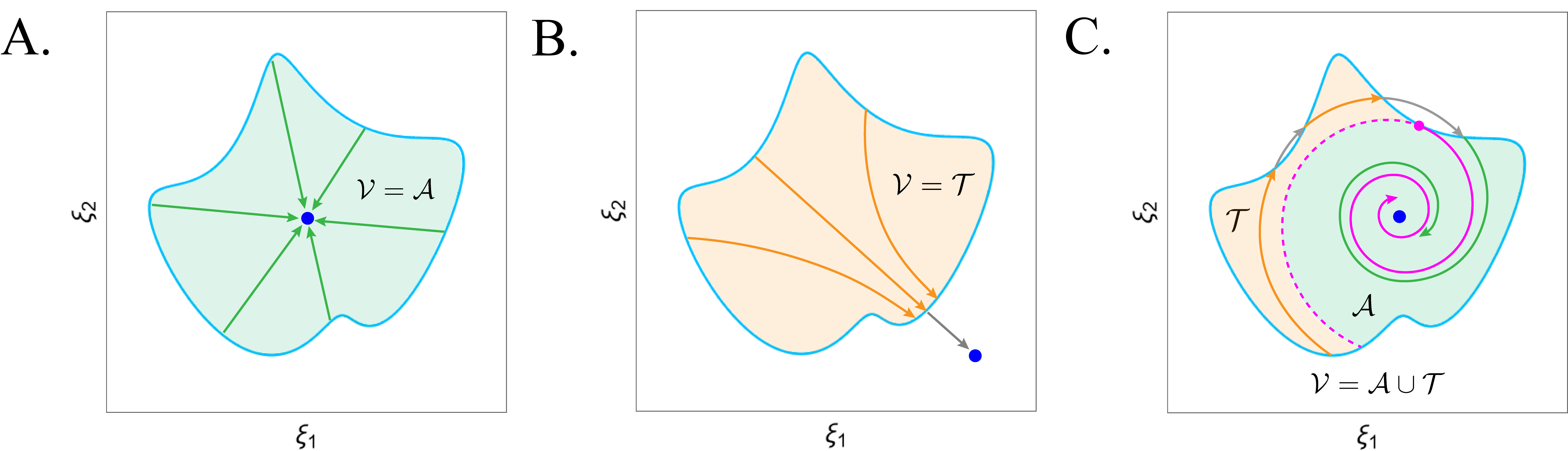}
\caption{\textbf{\textbf{Schematic of viability space decomposition for a single cell.} }
\textbf{A. }A single attractor (dark blue) is located within the viability region such that all of the vectors on the boundary point inward, and the entire region is \textit{asymptotically viable} (green). \textbf{B. }The attractor is outside of the viability region, such that every initial condition leads to a trajectory that will die in finite time and will never finish the path to the attractor (gray), making the whole region \textit{transiently viable} (orange). \textbf{C. }The attractor is within the boundary of the viability region, but some of the initial conditions in its would-be basin die in the transient. To decompose the viability region into its asymptotically and transiently viable sets, we find where the vector field is tangent to the viability boundary (magenta point). Since this point leads to an asymptotically viable trajectory, it is a \textit{mortality point}, and its backward time trajectory is a \textit{mortality manifold} that separates the two sets. }
\label{fig3}
\end{figure}

If an attractor’s would-be basin contains both trajectories that will live and die, how can we know which initial conditions are safe? To distinguish between these sets, viability space decomposition starts by looking for the points where the change vector is tangent to the boundary (magenta point). Assuming a smoothly varying segment of the viability boundary, any transition from vectors that result in immediate death to those that move into the interior will have to pass through such a tangent vector. If the forward-time trajectory of the tangent vector is asymptotically viable, then the backward-time trajectory can form a \textit{mortality manifold }(dashed magenta) that separates asymptotically and transiently viable sets.

While mortality manifolds paired with attractors and their basins are sufficient for carving out the asymptotically and transiently viable states, the multi-agent case demands more— we must ask which cells die, in what order, and what happens to those that survive. Doing this necessitates a hybrid dynamical system formalism that captures the continuous state evolution of cells within their viability constraints, and the instantaneous change that occurs when these constraints are violated \cite{aihara_theory_2010}. For a given group of interacting cells, we can represent this as a directed graph, where nodes are possible combinations of individuals and edges are death events (Fig. 3A). The most likely transitions will be those where a single cell dies, but intersections of different cells’ viability boundaries make it possible for multiple cells to die simultaneously. Importantly, these nodes and edges are only possible configurations and transitions, and which ones actually occur will depend on the dynamics. 

\begin{figure}
\centering
\includegraphics[width=4.6in]{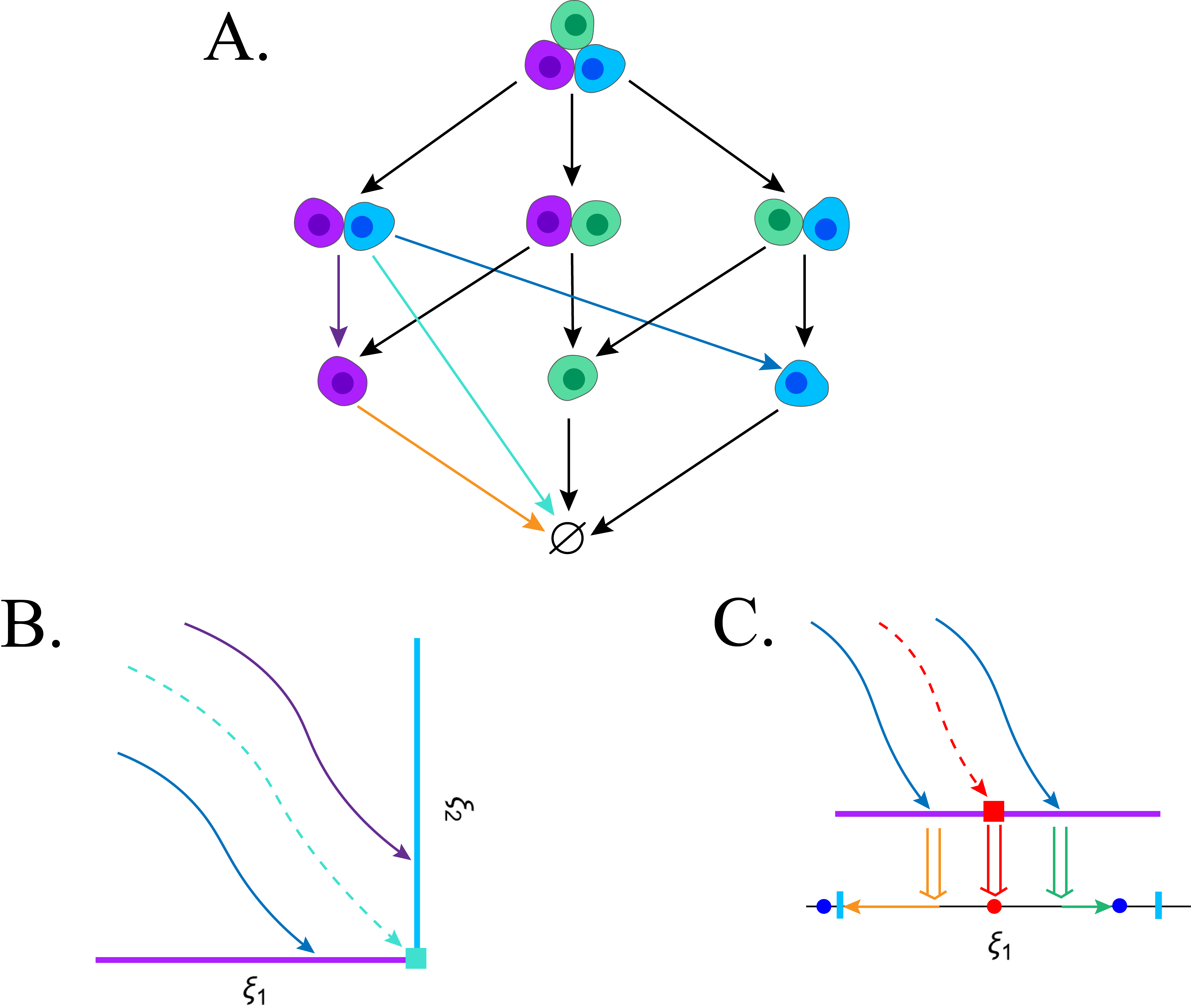}
\caption{\textbf{\textbf{Multicellular viability space decomposition as a hybrid dynamical system.}}
\textbf{A. }It is possible to visualize a multicellular model as a type of directed graph. Each node represents the cells currently alive with the corresponding continuous state space, and edges are the transitions that take place between nodes when one or more cells simultaneously die. Colored edges correspond to specific trajectories in the following subplots. \textbf{B. }Where both cells’ viability constraints meet (cyan square), the dynamics result in both cells perishing simultaneously. The backward time trajectory (dashed cyan trajectory) of this point forms an \textit{ordering manifold} where either the blue or purple cell will have its viability constraints violated first. \textbf{C. }When the purple cell dies, the fate of the blue cell depends on its own state as it collapses into its independent one-dimensional space (double arrows). If the blue cell falls to the left of its unstable equilibrium point (red dot), it will head towards a terminal attractor, and if it falls to the right, it will be asymptotically viable. To know which initial conditions in the joint blue-purple cell space will result in either outcome, we can find the terminal point that collapses onto the unstable equilibrium (red square) and integrate it backward in time to get a \textit{collapse manifold} (dashed red trajectory).}
\label{fig4}
\end{figure}

What new manifolds, in addition to mortality manifolds, will allow us to decompose the survival outcomes in this hybrid dynamical system? Consider Fig. 3B, which focuses on the survival outcomes when only the purple and blue cells are still alive. Even as the cells’ coupled configuration pushes both their essential variables toward their viability boundaries, which one dies first varies with the initial condition. At the intersection of the cells’ viability constraints (cyan square), we observe that the dynamics are immediately fatal for both. This rare state of joint fatality can provide us with global insight: its backward time trajectory creates an \textit{ordering manifold }that separates regions where either the blue or purple cell dies first. Paired with the basins of attraction and mortality manifolds, they reveal which cells in the immediate configuration will perish, but not what happens afterwards.

If we want to know what happens after a death event, we need to explore the dynamics of those cells still alive. In our example, Fig. 4C shows how, when the purple cell dies, the blue cell collapses into different parts of a lower-dimensional space depending on its state. This new state space, belonging to the blue cell alone, has both a viable and nonviable attractor, separated by an unstable equilibrium (red dot). If we want to know which initial conditions in the joint cell space will lead to the blue cell falling on either side of this basin boundary, we can look for which one of the fatal vectors in the two-cell system results in the blue cell falling exactly on this unstable node (red square). We can then integrate backward in time from this fatal point to find the manifold that separates these outcomes, called a \textit{collapse manifold}. Note that \textit{any }manifold that organizes a lower-dimensional subspace can be extended backward in higher-dimensional spaces as a collapse manifold, making them incredibly common in multicellular systems. While we have demonstrated these manifolds in low-dimensional schematics, in principle they mathematically scale as \(n-1\)-dimensional manifolds, much like the viability boundaries in the previous section. Actually deriving and visualizing these manifolds in high-dimensional state spaces is a challenge in-and-of itself, but even knowing of their existence can enable us to analyze cell models in a more strategic manner.

It is worth noting that viability space decomposition is only one proposal for addressing viability in a particular class of models. Other approaches have been developed from a control-theoretic perspective for both cells \cite{himeoka_theoretical_2024} and machines more broadly \cite{aubin_viability_2011}, and information-theoretic measures have also been proposed \cite{egbert_methods_2018, kolchinsky_semantic_2018,bartlett_physics_2025}. Despite this diversity, all these frameworks, including viability space decomposition, begin by assuming a known set of living states. In the next section, we discuss how we might begin to understand the principles that give rise to the viability regions themselves.

 \section{Origins of the Life-Death Boundary}

So far, we have focused on the principled decomposition of viability regions based on the dynamics that play out within them, but where does the life-death boundary come from in the first place? What determines its shape and structure? Can it be derived from physical and chemical principles? If we are to develop a more fundamental understanding of cell death proper, these features must be understood in a theoretically rigorous way. To have this conversation, we must first delineate between the \textit{extrinsic} and \textit{intrinsic} perspectives of viability constraints \cite{mcshaffrey_decomposing_2023, beer_deriving_2024}.

The examples we have explored to this point look at \textit{extrinsic viability}, in which the modeler explicitly defines and enforces the boundary that separates life from death within the cellular model. This approach is riddled with highly idealized and subtle assumptions about when a cell should be considered dead, but it is often necessary to have such an operational definition for survival. Perhaps surprisingly, extrinsic viability is also the dominant language of experimental approaches to cell death – we only measure indicators of death rather than death itself. In the case of necrosis, there is an ongoing debate of when in the sequence of metabolic collapse, oncosis (rapid swelling due to osmosis), lysis, and calcification death actually occurs \cite{majno_apoptosis_1995,trump_pathways_1997,ghaffarizadeh_physicell_2018}. Similarly, sporulation allows cells to remain dormant under harsh conditions with near nonexistent metabolic activity, making it hard to gauge whether they are still viable (Lennon \& Jones, 2011). In a recent interview, cell biologist Shai Shaham noted that sometimes it is as if the only way to know whether a cell has died is to make sure it is gone in its entirety \cite{strogatz_how_2024}, but this approach can just as easily significantly overshoot the time of death. For most practical purposes, the best we can often do is be explicit in what we approximate the limits of viability to be.

Unlike its extrinsic counterpart, the perspective of \textit{intrinsic viability} views cells as precarious systems of active matter that emerge from, and are capable of disintegrating back into, the underlying medium \cite{shaebani_computational_2020,beer_theoretical_2023}. In this case, the physiological limits of cells are not imposed by a modeler, but are realized by the limits of the interconnected network of processes needed to continuously regenerate the cell \cite{maturana_autopoiesis_1980, varela_principles_2025}. While this may sound abstract, there is evidence for it being a promising way to think about cellular persistence. In yeast, it has been suggested that the transition from dormancy to death occurs as the cells lose the ability to express genes, halting processes vital to survival \cite{maire_dormancydeath_2020}. Similarly, while it was previously thought that the outer membrane of mitochondria becoming porous was a commitment to apoptosis, more recent work demonstrates that cells can recover as long as enough mitochondria survive to stave off metabolic collapse and repopulate the cell \cite{spencer_measuring_2011, bock_mitochondria_2020}. 

While it would be ideal to exhaustively interrogate a cell's emergent viability constraints at the level of its physical or chemical processes, we are currently very far from this capability. Even if we had the technological capacity to simulate a whole cell at the level of underlying matter, it is not immediately clear how we would go about analyzing its viability in a principled way. At what granularity should we define a process, and which processes are fundamental to the cellular unit? Even gene expression stands on uncertain grounds, as neutrophils will continue to be transiently active after they have repurposed and expelled their DNA as an extracellular trap \cite{desai_matters_2016}. In all likelihood, a cell’s viability depends on the coordinated states of numerous processes spanning spatiotemporal scales such that any curated subset will never be universally predictive of immediate survival.

Despite these challenges, it is possible for us to begin at least conceptually exploring the nature of intrinsic viability in idealized models. Throughout the physical and life sciences, idealized or “toy” models have proven themselves as a means to build our intuitions about phenomena that initially seemed otherwise intractable \cite{beer_animals_2009,beer_milking_2024}. We propose that a similar route is available for studying intrinsically generated viability constraints. Recently, it was demonstrated that it is possible to derive the intrinsic viability limits of metastable patterns in Conway’s Game of Life (GoL), with applications for other cellular automaton models \cite{beer_deriving_2024}. While cellular automata may seem far from real biology, they can be seen as realizing a simple, spatially extended artificial chemistry \cite{dittrich_artificial_2001,beer_characterizing_2015}. For example, grid locations in GoL have a binary state space and will turn “on” and “off” according to the sum of “on” cells in their Moore neighborhood, resulting in processes of production, destruction, and maintenance. More recently, Beer demonstrated that we can also characterize \textit{partial processes}, which appear when some of the surrounding grid locations have unknown states \cite{beer_integrated_2020}. With these methods, it becomes possible to characterize the network that continuously regenerates the glider, as well as the limits at which this network breaks down \cite{beer_integrated_2020,beer_deriving_2024}. By looking at how this network interfaces with the glider’s local environment, we can partition the set of all glider-environment configurations based on whether they enable a transition that maintains the glider (interior of the viability region) or the disintegration of it (boundary of the viability region). Whereas extrinsic viability defines the viability boundary entirely separate of the dynamics, intrinsic viability shows that states belong to the boundary precisely because their transition results in the instantaneous loss of the individual. In Fig. 5, we show an example of the glider undergoing a sequence of viable transitions based on its local interactions with its environment and a final transition at the viability boundary where it disintegrates.  

\begin{figure}
\centering
\includegraphics[width=4.2in]{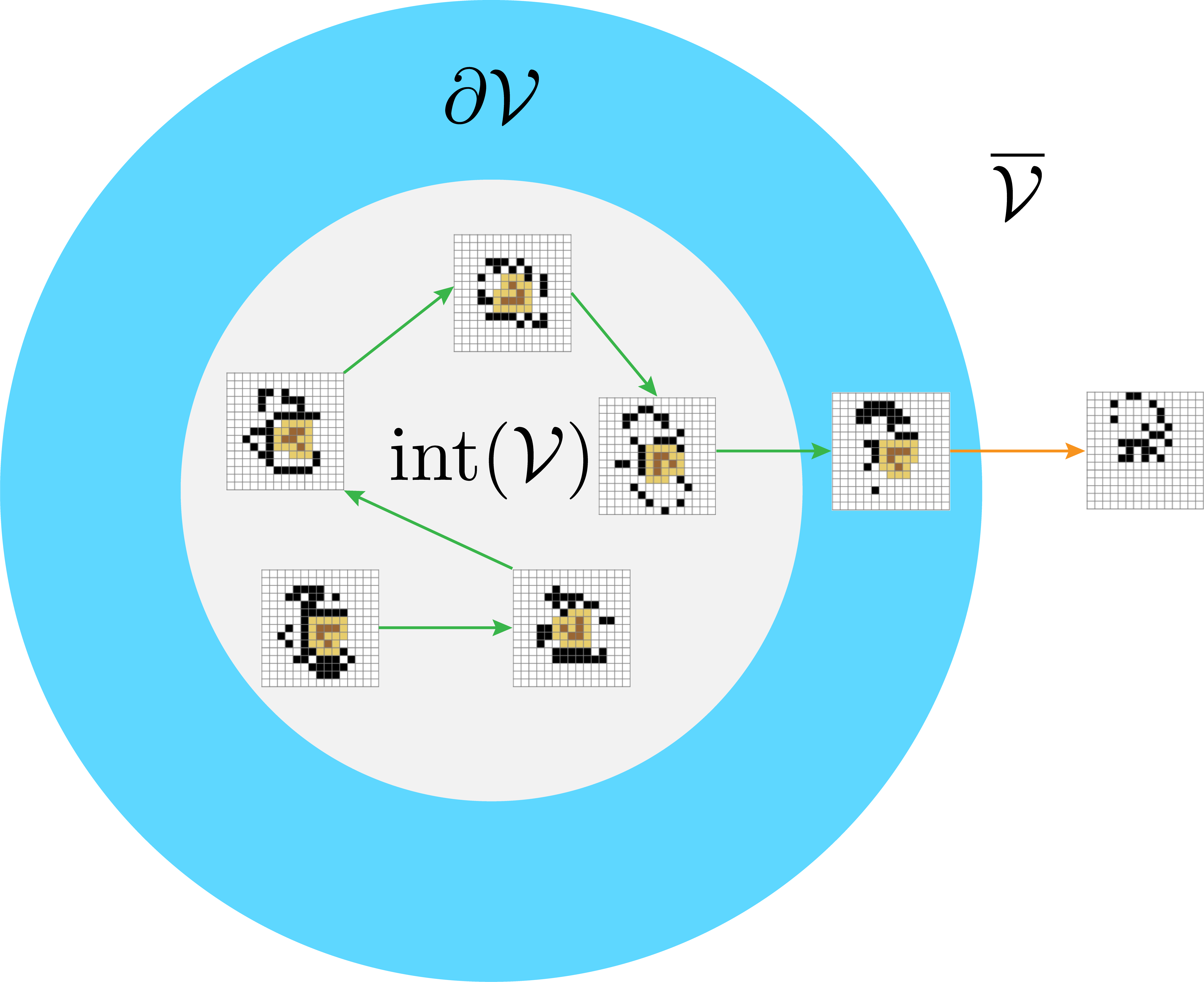}
\caption{\textbf{\textbf{A schematic of the intrinsic viability region of a glider in Conway’s Game of Life.}}
The glider is defined by a pattern of on-cells (brown) and a surrounding layer of off-cells (yellow) that function as its physical boundary or membrane. As the glider moves through its environment, it encounters various environmental configurations that function as perturbations, which will either result in a viable (green) or nonviable transformation (orange) according to its conditions for self-maintenance. The states that lead to the green transitions are in the interior of the viability region (light gray), and the state that results in a terminal transition belongs to its boundary (light blue). The complement of the viable set are all the configurations that do not contain the glider.}
\label{fig5}
\end{figure}

Starting from these simple models where intrinsic viability can be rigorously defined, we can incrementally build towards more realistic representations of living systems by systematically introducing new layers of complexity (Fig. 6). Already, steps have been taken towards generalizing this derivation to the Larger than Life cellular automaton family, which has a Euclidean limit \cite{gaul_autopoiesis_2024}. From there, a next step towards biophysical realism is to derive the intrinsic viability constraints of emergent protocells in reaction-diffusion systems \cite{agmon_exploring_2016,agmon_structure_2016}, where self-maintaining chemical networks begin to take on spatial and morphological organization. A central challenge for this path is defining the continuous generalization processes and the dependencies between them. Alternatively, one can pursue a complementary path by preserving molecular discreteness and developing lattice-gas or particle-based models with chemical reactions \cite{dab_lattice-gas_1991,rothman_lattice-gas_1994}. These models provide a natural bridge to mesoscale representations of metabolism and transport, where our current understanding of discrete biological processes might be more easily applied. Ultimately, these routes ought to converge toward the goal of constructing \textit{whole-cell models} whose viability constraints arise intrinsically from the coupling of molecular, spatial, and energetic processes. 

\begin{figure}
\centering
\includegraphics[width=5.6in]{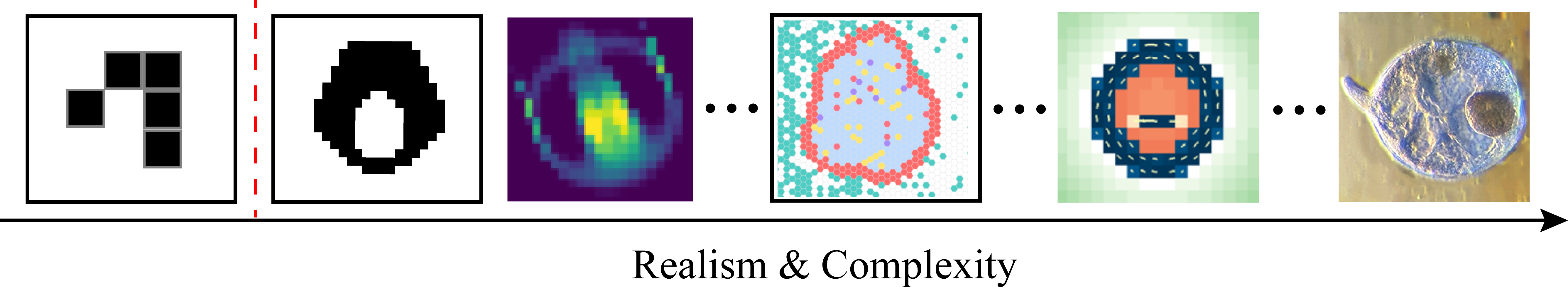}
\caption{\textbf{\textbf{Scaling the complexity of idealized models with intrinsic viability constraints.}}
We can begin to understand intrinsic viability by looking at idealized models that contain metastable systems that have the capacity of self-maintenance. The red dashed line marks the boundary of what we are currently capable of understanding (cellular automata like GoL), and beyond are more complex systems that slowly become more biologically realistic. Left to right: Conway’s Game of Life, Larger than Life cellular automata \cite{evans_larger_2001}, Lenia \cite{chan_lenia_2018, chan_lenia_2020}, a schematic of a particle-based (potentially Lattice Gas) model as generated by Claude Opus 4.1 \cite{dab_lattice-gas_1991,rothman_lattice-gas_1994}, reaction-diffusion models \cite{agmon_exploring_2016,agmon_structure_2016}, a real cell (of the \textit{Phacus} genus).}
\label{fig6}
\end{figure}

\section{Discussion}

In this paper, we examined how death is represented in contemporary computational models of cells, from subcellular networks to multicellular systems. We found  that many models do not explicitly define what conditions they consider sufficient for death, and those that do rely on ad hoc or externally imposed viability constraints. Even when death conditions are explicitly defined, they are often treated as secondary to the model’s dynamics, limiting our ability to understand how these constraints shape and interact with living processes.

To begin remedying these issues, we argued that viability constraints must be treated as a first-class concept in computational cell biology.  By framing life and death geometrically, we introduced viability space decomposition as a means  to partition viability regions into subspaces of qualitatively distinct survival outcomes – structured by classical separatrices and new organizing manifolds such as mortality, ordering, and collapse manifolds. More broadly, we have reframed existing models – whether ODE, agent-based, or multiscale frameworks – as instances of extrinsic viability, where survival limits are imposed by the modeler rather than emerging from the system itself. By developing simplified idealized models that exhibit intrinsic viability, we begin to bridge the conceptual gap between imposed and emergent life-death boundaries.

What does the future of viability look like? One active area of research for viability space decomposition involves  identifying bifurcations where mortality, ordering, and collapse manifolds appear and disappear, alongside corresponding regions with particular survival outcomes. Recent research has also shown that viability space decomposition can still offer qualitative insights when dynamics include moderate additive noise \cite{mcshaffrey_shaking_2025}; however, further research is needed to develop analysis techniques for other types of stochastic dynamics and less rigid viability constraints. Developing a taxonomy of models and their associated viability constraints, along with analytical or computational tools suited to each, will help formalize this emerging framework. Recent multimodal single-cell analyses \cite{hao_integrated_2021,weiler_cellrank_2024} also reveal empirical manifolds of cell-state transitions that resemble these theoretical dynamical trajectories, and it may be productive to align viability space manifolds with these geometries.

In the context of whole-cell models, a central challenge is how to compose and reconcile conflicting viability constraints across submodels, integrating metabolic, genetic, and spatial processes into a coherent viability boundary. It would also be interesting to construct something analogous to the directed graph in multicellular viability space decomposition, except that the integrity of submodels would be presented instead of distinct cells. In the case of multicellular whole-cell models, the graph structure would be more complicated with additional nested structure. These developments connect directly to biomedical digital twins \cite{casola_opportunities_2023}, where our ability to analyze viability outcomes is of central importance.

Beyond these advances in extrinsic viability, it is equally important to deepen our understanding of intrinsically generated viability constraints. A key next step is to approximate the viability constraints of high-dimensional, emergent cellular systems. Establishing such approximations would not only strengthen our conceptual grounding of intrinsic viability but also create a bridge between theoretical models and clinically relevant digital twins, enabling simulations that capture how living systems traverse, resist, or succumb to the boundaries that delimit life itself. Ultimately, a mature theory of viability would unify the extrinsic and intrinsic perspectives  – linking the constraints used in whole-cell and multicellular models with the intrinsic limits revealed in idealized model systems. 

\section*{Acknowledgments}

This material is based upon work supported by the National Science Foundation Graduate Research Fellowship Program under Grant No. 2240777. Any opinions, findings, and conclusions or recommendations expressed in this material are those of the author(s) and do not necessarily reflect the views of the National Science Foundation. E.A. is funded by NSF award OCE-2019589 to the Center for Chemical Currencies of a Microbial Planet, by NIH award P41GM109824 to the Center for Reproducible Biomedical Modeling, and John Templeton Foundation award \#62825.

\bibliographystyle{abbrv}  
\bibliography{life_death_references}

\end{document}